\documentclass[11pt, a4paper,nobibnotes, %nofootinbib
]{article}
\usepackage{amssymb,eurosym,amsfonts,amsmath,calc}
\usepackage{array}
\usepackage{indentfirst}
\usepackage{graphicx}
\usepackage{hyperref}
\usepackage{authblk}
\usepackage{color}
\usepackage{multicol}
\usepackage[utf8]{inputenc}
\usepackage[normalem]{ulem}

\begin{document}
\title{$\mathcal{N}=2$ resonant superalgebra for supergravity}
	\author{Remigiusz Durka \thanks{remigiusz.durka@uwr.edu.pl}}
	\author{Krzysztof M. Graczyk \thanks{krzysztof.graczyk@uwr.edu.pl}}
\affil{University of Wroc\l{}aw, Institute for Theoretical Physics, pl.\ M.\ Borna 9, 50-204 Wroc\l{}aw, Poland}
\setcounter{Maxaffil}{0}
\renewcommand\Affilfont{\itshape\small}

\maketitle

%%%%%%%%%%%%%%%%%%%%%%%%%%%%%%%%%%%%%%%%%%%%%%%%%%%%%%%%%%%%%%%%%%%%%%%%%%%%%%%%%%%%%%%%%%%%%

\begin{abstract}
	We present new superalgebra for $\mathcal{N}=2$ $D=3,4$ supergravity theory endowed with the $U(1)$ generator. The superalgebra is rooted in the so-called Soroka-Soroka algebra and spanned by the Lorentz $J_{ab}$ and Lorentz-like $Z_{ab}$, translation $P_a$ and $T$ generators, as well as two supercharges $Q^I_\alpha$. It is the only possible realization for a given generator content. We construct a corresponding 3D Chern-Simons supergravity realization of the superalgebra and discuss its relevance.     
\end{abstract}

%%%%%%%%%%%%%%%%%%%%%%%%%%%%%%%%%%%%%%%%%%%%%%%%%%%%%%%%%%%%%%%%%%%%%%%%%%%%%%%%%%%%%%%%%%%%%

\section{Introduction}

Supergravity incorporates the interchangeability of the fermionic and bosonic fields under the so-called supersymmetry. Consequently, the graviton (spin-$2$) comes with its superpartner gravitino (spin-$3/2$). A natural generalization is obtained in $\mathcal{N}$-extended supergravity model containing $\mathcal{N}$ gravitinos. Obviously $\mathcal{N} = 0$ gives  general relativity while $\mathcal{N} = 1$ corresponds to standard supergravity \cite{Freedman:1976xh, Deser:1976eh, Townsend:1977qa}. 

An effort has been made to construct the unified theory of gravity with other interactions, e.g., such as electromagnetism. In the latter, the unification is realized within an algebraic framework of supergravity. An example of such theory is $\mathcal{N} = 2$ supergravity~\cite{Ferrara:1976fu}. In this model, the graviton (spin-$2$) and gravitino (spin-$3/2$) gauge action couples with another gravitino (spin-$3/2$) and spin-$1$ vector field. Note that the action is obtained from the supersymmetry group, which includes the central charges~\cite{Haag:1974qh} and rotation of two gravitinos~\cite{VanNieuwenhuizen:1981ae}.

We distinguish two basic initial algebraic structures for supergravity construction, namely, Poincar\'e and AdS (Anti de-Sitter) algebras. In this paper we do not consider the Poincar\'e scenario \cite{Ferrara:1976fu, Howe:1995zm} but we focus on the AdS $\mathcal{N}=2$ case \cite{Townsend:1977fz, Ortin:2015hya, Gocanin:2019ota}. The corresponding superalgebra in $D=4$ with the Lorentz generator $J$, translation $P$, $U(1)$ generator $T$, and two supercharge~$Q^I$ generators is expressed as:
\begin{align}\label{superalgebra_AdS}
	\left[ J_{ab},J_{cd}\right] &=\eta_{bc}J_{ad}-\eta_{ac}J_{bd}+\eta_{ad}J_{bc}-\eta_{bd}J_{ac}\,,\notag\\
	\left[ J_{ab},P_{b}\right] &=\eta _{bc}P_{a}-\eta _{ac}P_{b}\,,\qquad\left[ P_{a},P_{b}\right] =J_{ab}\,,  \notag\\
	\left[J_{ab},Q^I_{\alpha }\right] &=\frac{1}{2}\,\left( \Gamma _{ab}\right) _{\alpha}^{\beta }Q^I_{\beta }\,,\qquad \left[ P_{a},Q^I_{\alpha }\right]	=\frac{1}{2}\,\left( \Gamma _{a}\right) _{\alpha }^{\beta }Q^I_{\beta}\,,\qquad \left[T,Q^I_{\alpha}\right]=\epsilon^{IJ} Q^J\,,\notag\\ \left\{ Q^I_{\alpha },Q^J_{\beta }\right\} &=-\delta^{IJ}\left(\left( \Gamma^{a} C\right) _{\alpha \beta } P_{a} -\frac{1}{2} ( \Gamma ^{ab}C )_{\alpha \beta} J_{ab} \right)+\epsilon^{IJ}C_{\alpha\beta}T\,.
\end{align} 
By $\alpha,\beta$ we denote spinorial indices, whereas $a,b=0,1,2,3$ are the Lorentz indices, and $I=1,2$. By $\eta_{ab}=\mathrm{diag}(-,+,+,+)$ we denote flat Minkowski metric, and by $C_{\alpha\beta}$ the anti-symmetric charge conjugation matrix. 

By gauging the superalgebra, one can define the connection, curvature, corresponding gauge transformations, and Killing forms. With these definitions, we can proceed with constructing the action being invariant under Lorentz and supersymmetry transformations.

The Poincar\'e $\mathcal{N}=2$ supergravity contains $SO(2)$ automorphisms group generated by $T$. Generator $T$ rotates supercharges. Simultaneously there is another $U(1)$ generator playing the role of the central charge in the anticommutator of $\{Q^I_\alpha, Q^J_\beta\}$. The Killing form between them is non-zero \cite{Howe:1995zm}, leading to a mixed $U(1)$ term in the gauge action. While in the AdS case, we have a single generator $T$ in both mentioned the (anti)-commutation rules. Hence we can not call it a central charge. However, this way, we achieve the non-zero Killing form $\langle T\, T\rangle$ and a standard $U(1)$ term in the gauge action, which is not possible in the Poincar\'e theory.

The $\mathcal{N}$-extended theories are obtained by enlarging the number of fermionic generators. A similar mechanism has been used for the bosonic counterpart in recent years. It is achieved by including additional Lorentz-like and translational-like generators to the base of Poincar\'e/AdS algebras. This research direction takes its origin from the so-called Maxwell algebra~\cite{Schrader:1972zd, Bacry:1970ye}, and further development obtained by the so-called Semigroup expansion (\textit{S-expansion}) procedure~\cite{Izaurieta:2006zz, Salgado:2013eut, Andrianopoli:2013ooa, Concha:2016hbt, Concha:2016kdz}. It provides a consistent way of deriving new (super)algebras. The essence of the semigroup expansion method comes down to maintaining the same pattern of structure constants in the (anti)-commutation rules for the newly added generators as for primary (supersymmetric) AdS algebra. The Soroka-Soroka \cite{Soroka:2006aj} or Maxwell algebras \cite{Schrader:1972zd} are natural enlargements of AdS and Poincar\'e algebras, respectively, but there are merely two examples along the way \cite{Salgado:2014jka,Salgado-Rebolledo:2019kft}. Our past work \cite{Durka:2021glk} explored the question of all (super)algebra realizations. By going beyond the generating new algebras by the semigroup rescaling and focusing purely on generators and fulfilling the Jacobi identities, as well as including zeros in the commutation outcomes, we arrive at explicit algebras we call ``resonant``.

The resonant/semigroup algebras were analyzed in the context of various applications: the context of the cosmological constant term \cite{deAzcarraga:2010sw}, non-relativistic symmetries~\cite{Concha:2019mxx, Concha:2019lhn, Concha:2020eam, Concha:2020tqx}, as well as asymptotic symmetries of 3D CS \cite{Concha:2018zeb, Concha:2018jjj}. Some attempts to obtain 3D solutions can be found in \cite{Concha:2018zeb} and \cite{Concha:2018jjj}. Additionally, it is worth mentioning a bi-gravity configuration in 3D~\cite{Hoseinzadeh:2017lkh}, which leads to some preliminary notions of bi-supergravity in our last paper~\cite{Durka:2021glk}. We also mention the ansatz and the "Maxwelian" symmetry used in the 4D cosmology configuration~\cite{Durka:2011va}, as well as 3D topological insulators within Hall effect framework~\cite{Palumbo:2016nku, Durka:2019guk}, where the standard $U(1)$ field is immersed within Maxwellian field associated with the $Z_{ab}$ generator. However, still, the physical interpretation of the new fields induced by new generators remains unclear.

The discussion of all algebraic possibilities for a given set of generators in the bosonic sector was completed in papers~\cite{Durka:2016eun, Durka:2019vnb}. After a series of particular examples \cite{Soroka:2006aj, Bonanos:2009wy, Bonanos:2010fw, Kamimura:2011mq, deAzcarraga:2014jpa, Durka:2011gm, Durka:2012wd, deAzcarraga:2012zv, Fierro:2014lka, Penafiel:2018vpe, Andrianopoli:2021rdk, Concha:2014tca} of supersymmetric versions, the resonant $\mathcal{N}=1$ cases were concluded in~Ref.~\cite{Concha:2019icz}. In our recent paper~\cite{Durka:2021glk}, after an extensive search, we provided the complete list of resonant superalgebras up to two fermionic generators.

In the present paper, similarly to AdS scenario, we investigate the $\mathcal{N}=2$ framework with $T$ generator while still considering the resonant enlargement on the bosonic counterpart. Using the method described in Ref.~\cite{Durka:2021glk}, we consider the algebra structure, $JPZ+QY+T$, that contains the Lorentz $J_{ab}$, translation $P_{a}$ and the additional Lorentz-like generator $Z_{ab}$. Note that this superalgebra contains two fermionic Majorana supercharges $Q_{\alpha}^I$, ($Q_{\alpha}^1\equiv Q_{\alpha}$ and $Q_{\alpha}^2 \equiv Y_{\alpha}$), and the $U(1)$ generator, denoted as $T$. Interestingly, we have obtained only one possible realization of the superalgebra in such a case.
Quite recently, in Ref. \cite{Concha:2020tqx} the $\mathcal{N}=2$ algebra, similar to our findings, has been discussed. However, it represents a distinct scenario because it is defined by a larger number of generators, which eventually leads to different invariant tensors and, as a result, modification of the gauge action.

In further sections, we will construct the 3D Chern-Simons (CS) Lagrangian \cite{Achucarro:1986uwr, Hassaine:2016amq, Concha:2015woa} based on the 3D version of the derived superalgebra. We discuss the differences between the newly obtained supergravity model and the previous AdS result. 
Similar studies can be found in literature, \cite{Bonanos:2009wy, Concha:2019icz}. However, we provide a much simpler algebraic structure, which originates from the so-called Soroka-Soroka enlargement \cite{Salgado:2013eut, Soroka:2006aj, Soroka:2011tc} instead of the one based on Maxwell algebra \cite{Schrader:1972zd, Bacry:1970ye}. In our approach, we avoided considering additional bosonic $U(1)$ generators. 

The paper is organized as follows: in Sec.~\ref{Sec:RESALG} we give new $\mathcal{N}=2$ superalgebra; in Sec.~\ref{Sec:Lagrangian} we construct the respective action for the 3D Chern-Simons model. We complete the paper with the summary and outlook in Sec.~\ref{Sec:Outlook}. The paper also contains three appendixes. 

%%%%%%%%%%%%%%%%%%%%%%%%%%%%%%%%%%%%%%%%%%%%%%%%%%%%%%%%%%%%%%%%%%%%%%%%%%%%

\section{$\mathcal{N} = 2$ Soroka-Soroka superalgebra}
\label{Sec:RESALG}

The main idea of our research program is to perform a systematic search for the algebra structures suitable for supergravity models with an arbitrary number of generators of certain types. Following the ``resonant`` method introduced in Ref.~\cite{Durka:2021glk}, we consider, besides Lorentz and translations generators, also bosonic Lorentz-like $Z_{ab}$. We include the $T$ generator, incorporating the $U(1)$ field (associated with $T$) into the supergravity framework. The obtained results are valid for 3D and 4D; for more explanation, see \cite{Durka:2021glk}. 

The structure of superalgebras with $T$ generator is more complicated than the ones without it. Similarly, as in Ref.~\cite{Durka:2021glk}, to find the superalgebras, we consider computer-generated thousands of algebra candidates ($14580$). The successful superalgebra must satisfy the Jacobi identities, therefore the Jacobi identities have to be computed and verified for each candidate. 

Although there are six $JPZ$ bosonic algebras altogether \cite{Durka:2021glk}, ultimately, we arrive at one distinct realization of $JPZ+QY+T$ superalgebra, which we shall discuss below. It is based on the Soroka-Soroka $\mathfrak{C}_4$ algebra \cite{Soroka:2006aj, Soroka:2011tc} and has the following form:
\begin{align}\label{resonant_algebra_JPZ+QY+T}
	\left[ J_{ab},J_{cd}\right] &=\eta_{bc}J_{ad}-\eta_{ac}J_{bd}+\eta_{ad}J_{bc}-\eta_{bd}J_{ac}\,,\notag\\
	\left[ J_{ab},Z_{cd}\right] &=\eta_{bc}Z_{ad}-\eta_{ac}Z_{bd}+\eta_{ad}Z_{bc}-\eta_{bd}Z_{ac}\,,\notag\\
	\left[ Z_{ab},Z_{cd}\right] &=\eta_{bc}Z_{ad}-\eta_{ac}Z_{bd}+\eta_{ad}Z_{bc}-\eta_{bd}Z_{ac}\,,\notag\\
	\left[ J_{ab},P_{b}\right] &=\eta _{bc}P_{a}-\eta _{ac}P_{b}\,,\notag\\
	\left[ Z_{ab},P_{b}\right] &=\eta _{bc}P_{a}-\eta _{ac}P_{b}\,,\notag\\
	\left[ P_{a},P_{b}\right]&=Z_{ab}\,,\notag\\
	\left[J_{ab},Q^I_{\alpha }\right] &=\frac{1}{2}\,\left( \Gamma _{ab}\right) _{\alpha}^{\beta }Q^I_{\beta }\,,\qquad \left[ P_{a},Q^I_{\alpha }\right]	=\frac{1}{2}\,\left( \Gamma _{a}\right) _{\alpha }^{\beta }Q^I_{\beta}\,,\notag\\
	\left[ Z_{ab},Q^I_{\alpha }\right]&=\frac{1}{2}\,\left( \Gamma _{ab}\right) _{\alpha }^{\beta }Q^I_{\beta}\,,\qquad \left[T,Q^I_{\alpha}\right]=\epsilon^{IJ} Q^J\,\notag\\
	\left\{ Q^I_{\alpha },Q^J_{\beta }\right\} &=-\delta^{IJ}\left(\left( \Gamma^{a} C\right) _{\alpha \beta } P_{a} -\frac{1}{2} ( \Gamma ^{ab}C )_{\alpha \beta} Z_{ab} \right)+\epsilon^{IJ}C_{\alpha\beta}T\,.
\end{align}

Derivation was based on the consistency with commutation relations $[T, Q^I]=\epsilon^{IJ} Q^J$ (rotating $[T, Q]\sim Y$ and $[T, Y]\sim Q$), commutator $[J,X] \sim X$ (with $X$ being any generator, in particular, $[J, Q^I]\sim Q^I$), and the necessity of non-vanishing anticommutator $\{Q, Q\}\neq 0$, as well as $T$ appearing in the anticommutator $\{Q, Y\}\sim T$. Note that Soroka-Soroka $\mathfrak{C}_4$ algebra can be expressed as the direct sum of the AdS and Lorentz algebras \cite{Soroka:2006aj, Durka:2016eun}, which is achieved by an appropriate change of the basis. However, the resulting generators lose their physical role; see comment in Appendix~\ref{Appendix_C}.

We remind that there are six available bosonic $JPZ$ algebras altogether \cite{Durka:2021glk}, where besides the Soroka-Soroka algebra \cite{Soroka:2006aj, Soroka:2011tc} (also called \textit{AdS-Maxwell} or \textit{AdS-Lorentz}), we have the Maxwell algebra $\mathfrak{B}_4$ \cite{Schrader:1972zd, Bacry:1970ye} and four Poincar\'e-like algebras \cite{Durka:2016eun}. The bosonic subalgebra for newly obtained example turned out to be solely given by the Soroka-Soroka algebra.

Some effort has been made to achieve algebra based on Maxwell $\mathfrak{B}_4$ \cite{Schrader:1972zd, Bacry:1970ye}. However, they have ended with a much more complicated setup with additional $U(1)$ generators, or a modification, which includes the new Lorentz-like generator $Z_{ab}$ in $\{Q_\alpha, Y_\beta\}$ \cite{Concha:2019icz}. Note that $\gamma_5 C_{\alpha\beta}$ and central charge $T_C$ could be included in $\{Q^I_\alpha,Q^J_\beta\}$ \cite{Bonanos:2009wy}, but we do not consider this possibility for direct clarity. In the future, we will discuss a much broader scope and incorporate the resonant framework within the Poincar\'e $\mathcal{N}=2$ superalgebra.

As we mentioned, another interesting example of the $\mathcal{N}=2$ supergravity is given in Ref.~\cite{Concha:2020tqx}. This algebra also has roots in the bosonic Soroka-Soroka subalgebra. However, the main goal of \cite{Concha:2020tqx} was to achieve the $\mathcal{N}=2$ Maxwell superalgebra through the Inonu-Wigner contraction. Effectively, the obtained algebra structure is richer by one additional internal symmetry generator and a central charge in contrast to our case. Having two bosonic internal symmetry generators was essential to assure non-degenerate invariant tensors. We found that the Inonu-Winger contraction applied to our $\mathcal{N}=2$ Soroka-Soroka superalgebra also leads to $\mathcal{N}=2$ Maxwell algebra but with degenerated invariant tensors. The main idea of our research program is to perform a systematic search for the algebra structures suitable for supergravity models with an arbitrary number of generators of certain types. Therefore, we intentionally consider minimal generator content to achieve $\mathcal{N}=2$ resonant superalgebra and $U(1)$ term in action.

%%%%%%%%%%%%%%%%%%%%%%%%%%%%%%%%%%%%%%%%%%%%%%%%%%%%%%%%%%%%%%%%%%%%%%%%%%%%%%%%%%%%%%%%%%%%%%%%%%%%

\section{Supergravity action in $D=3$}
\label{Sec:Lagrangian}

As the construction of the 3D Chern-Simons model is straightforward and assures the full symmetry invariance, we give the 3D CS model based on the new superalgebra. In this case, we transit to dual definitions of generators $J_{a}=\frac{1}{2}\epsilon_{a}{}^{bc}\,J_{bc}$, spin connection $\omega_{a}=\frac{1}{2}\epsilon_{a}{}^{bc}\,\omega_{bc}$, as well as definitions like Lorentz curvature $\mathcal{R}^a(\omega)=d\omega^a+\frac{1}{2}\epsilon^{abc}\omega_b\,\omega_c$. We define also covariant derivative, which acting on dreibein gives $D_{\omega}e^{a}=d e^{a}+\epsilon^{abc}\omega_{b}\,e_{c}$, and acting on the Majorana spinor gives $D_\omega \psi=d \psi +\frac{1}{2}\omega^a \Gamma_a \psi$.

The necessary elements to construct Lagrangian are the gauge connection one-form~$\mathbb{A}$; the super-curvature two-form~$\mathbb{F}$; the gauge parameter~$\Theta$, along with the gauge transformations; and the Killing form $\left\langle \dots \right\rangle$ given in the form of the invariant tensors.

The connection contains additional Lorentz-like generator $Z_{a}=\frac{1}{2}\epsilon_{a}{}^{bc}\,Z_{bc}$ with a corresponding field  $h_{a}=\frac{1}{2}\epsilon_{a}{}^{bc}\,h_{bc}$, and two gravitinos, $\psi$ associated with $Q$ and $\chi$ associated with $Y$,
\begin{equation}
	\label{connection-mathfrakC4_1and2}
	\mathbb{A}=\omega^{a}J_{a}+\frac{1}{\ell} e^{a}P_{a}+ h^{a}Z_{a}+\frac{1}{\sqrt{\ell}}\psi^{\alpha}Q_{\alpha}+\frac{1}{\sqrt{\ell}}\chi^{\alpha}Y_{\alpha}+\frac{1}{\ell} a T\,.
\end{equation}
We also introduce the $\ell$ parameter to make dimensions right.
The super-curvature two-form $\mathbb{F}$ is built straightforwardly from the $\mathbb{A}$ connection
\begin{equation}
	\label{curvature-mathfrakC4_1and2}
	\mathbb{F}=F^{a}J_{a}+\frac{1}{\ell} T^{a}P_{a}+H^{a}Z_{a}+\frac{1}{\sqrt{\ell}}\mathcal{F}^{\alpha}Q_{\alpha}+\frac{1}{\sqrt{\ell}}\mathcal{G}^{\alpha}Y_{\alpha} +\frac{1}{\ell} f(a)\,T\,.
\end{equation}
Indeed, for the superalgebra~\eqref{resonant_algebra_JPZ+QY+T} we have:
\begin{align}
	F^{a}& =\mathcal{R}^{a}(\omega)\,,  \notag \\
	H^{a}& =D_{\omega}h^{a}+\frac{1}{2}\epsilon^{abc}h_b h_c+\frac{1}{2\ell^2}\epsilon^{abc} e_b e_c+\frac{1}{2\ell}\left(\bar{\psi}\Gamma^{a}\psi+\bar{\chi}\Gamma^{a}\chi\right)\,,\notag\\
	T^{a}& =D_\omega e^{a}+\frac{1}{2}\left(\bar{\psi}\Gamma^{a}\psi+\bar{\chi}\Gamma ^{a}\chi \right) \,,  \notag \\
	\mathcal{F} & =\mathcal{D}_{\omega }\psi +\frac{1}{2\ell}\,e^{a}\Gamma_{a}\psi  +\frac{1}{2} \,h^{a}\Gamma_{a}\psi-\frac{1}{\ell}a\,\chi\,,  \notag\\
	\mathcal{G} & =\mathcal{D}_{\omega }\chi +\frac{1}{2\ell}e^{a}\Gamma _{a}\chi +\frac{1}{2}\,h^{a}\Gamma _{a}\chi+\frac{1}{\ell}a\,\psi \,,\notag\\
	f& =da-\frac{1}{2}(\psi^{\alpha}C_{\alpha\beta}\chi^{\beta} -\chi^\alpha C_{\alpha\beta}\psi^{\beta})=da-\frac{1}{2}(\bar{\psi}\chi -\bar{\chi}\psi) \,.
\end{align}

The general the three-dimensional Chern-Simons action reads
\begin{equation}
	\label{Action-mathfrakC4_1}
	I_{CS}=\frac{k}{4\pi }\int_{\mathcal{M}}\left\langle \mathbb{A} \wedge d\mathbb{A}+\frac{1}{3} \mathbb{A}\wedge [[\mathbb{A},\mathbb{A}]]\right\rangle \,,
\end{equation}
where $[[.,.]]$ is the generalized commutator/anticommutator, and $\left\langle \dots \right\rangle$ is Killing form being used to contract all the group indices. Depending on the particular algebra \cite{Durka:2021glk, Durka:2016eun} we calculate Killing forms as the direct consequence of the evaluation of identity with (anti)-commutation relations:
\begin{align}
	\langle[[X_i , X_j ]]\, X_k \rangle =\langle X_i \, [[X_j , X_k ]] \rangle\,,
\end{align}
where $X_i$ is any generator. As a result the Killing metric takes the form of the invariant tensor given for any combination of two generators. As for the bosonic part:
\begin{align}
	\left\langle J_{a}\,J_{b}\right\rangle &=\alpha_{0}\,\eta_{ab}\,,\\
	\left\langle J_{a}\,P_{b}\right\rangle =\left\langle Z_{a}\,P_{b}\right\rangle &=\alpha_{1}\,\eta_{ab}\,,\\
	\left\langle J_{a}\,Z_{b}\right\rangle =\left\langle Z_{a}\,Z_{b}\right\rangle=\left\langle P_{a}\,P_{b}\right\rangle &=\alpha_{2}\,\eta_{ab}\,.
\end{align}
For the fermionic part we have:
\begin{align}
	\left\langle Q_\alpha\,Q_\beta\right\rangle = \left\langle Y_\alpha\,Y_\beta\right\rangle=2(\alpha_{1} + \alpha_{2})\,C_{\alpha\beta}\,.
\end{align}
The most important here, although $[T,T]=0$, is the non-degenerate
\begin{align}
	\left\langle T\,T\right\rangle &=2(\alpha_{1} + \alpha_{2})\,,
\end{align}
allowing for the $ada$ term in the action.

Note that the resonant/semigroup framework introduces the sub-invariant sectors by the arbitrary valued $\alpha$'s constants in front of invariant tensors.

The $JPZ+QY+T$ superalgebra~\eqref{resonant_algebra_JPZ+QY+T} has the corresponding 3D Chern-Simons action:
\small
\begin{align}
	I&=\frac{k}{4\pi }\int \left[ \alpha _{0}\left(
	\omega ^{a}d\omega _{a}+\frac{1}{3}\epsilon^{abc}\omega _{a}\omega
	_{b}\omega _{c}\right) \right.  \notag \\
	&\left. +\alpha _{1}\left( \frac{2}{\ell}\mathcal{R}^{a}e_{a}+\frac{1}{3\ell^3}\epsilon
	^{abc}e_{a}e_{b}e_{c}+\frac{2}{\ell}e_{a}D_{\omega }h^{a}+\frac{1}{\ell}\epsilon ^{abc}e_{a}h_{b}h_{c}+\frac{2}{\ell}
	\bar{\psi}\mathcal{F}+\frac{2}{\ell}\bar{\chi}\mathcal{G}+ \frac{2}{\ell^2} a\,f(a)\right) \right.  \notag\\
	&\left. +\alpha _{2}\left(\frac{1}{\ell^2}e_{a}D_{\omega }e^{a}+2h_{a}\mathcal{R}
	^{a}+\frac{1}{\ell^2}\epsilon^{abc}e_{a}e_{b}h_{c}+h_{a}D_{\omega }h^{a}+\frac{1}{3}\epsilon ^{abc}h_{a}h_{b}h_{c}+\frac{2}{\ell}\bar{\psi}\mathcal{F}+\frac{2}{\ell}\bar{\chi}\mathcal{G}+\frac{2}{\ell^2} a\,f(a)\right) \right] \,.\notag\\ \label{Action-mathfrakC4_2}
\end{align}\normalsize

Similarly, as in previous works \cite{Durka:2021glk, Durka:2019guk,Concha:2019icz, Concha:2020atg} the action can be decomposed into several sub-invariant sectors defined by dimensionless $\alpha_i$'s constants. The separation of terms into the sectors directly depends on the superalgebra. 

Let us underline that our model's supergravity action ($\alpha_1$ sector)   
is very similar to the AdS one. Indeed, the fermionic contributions are the same. A difference is induced by the presence of the field $h^{a}$ corresponding to the $Z$ generator.

In comparison to the $\mathcal{N}=2$ AdS case (see Appendix \ref{Appendix_B}), due to the generator $Z$, we have a new sector governed by the $\alpha_2$ constant. Moreover, the torsion contributes to $\alpha_2$ sector in the case of new algebra, whereas for the AdS, it appears in the ``exotic`` $\alpha_0$ sector. Hence, the full new supergravity model leads to modification of the fermionic sector.

Besides the dynamical $\alpha_1$ sector, our new action contains additional fermionic terms. The fermionic terms contribute to $\alpha_2$ sector, whereas $\alpha_0$ sector contains purely $CS(\omega)$.

We note that besides the term $a\,(\bar{\psi}\chi -\bar{\chi}\psi)$, resulting directly from a sum of $a\,f(a)$ and $\bar{\psi}\mathcal{F}+\bar{\chi}\mathcal{G}$ part, we got cancellations in the actions of other mixed fermionic terms due to $\langle Q_\alpha\,Y_\beta\rangle=0$. As a result, in the final action we have pure $U(1)$ term $\frac{2}{\ell^2} a\,da$ plus the mentioned $\frac{1}{\ell^2}a\,(\bar{\psi}\chi -\bar{\chi}\psi)$ contribution.

Comparing our action with the one from \cite{Concha:2020tqx}, we see that using a single internal symmetry generator (denoted in our paper by $T$) brings noticeable changes in the invariant tensors, which are reflected in the final action. In particular, in \cite{Concha:2020tqx}, we see the lack of "proper" $U(1)$ term in action's gravitational $\alpha_1$ sector. Indeed, $u d u$ term is present only in the additional $\alpha_2$ sector, while in our case, the term, denoted as $a d a$, is present in both $\alpha_1$ and $\alpha_2$ sectors.

%%%%%%%%%%%%%%%%%%%%%%%%% Outlook %%%%%%%%%%%%%%%%%%%%%%%%%%%%%%%%%%%%%%%%%%%%%%%%%%%%%%%%%%%%

\section{Outlook}
\label{Sec:Outlook}

We have introduced a new $\mathcal{N}=2$ superalgebra rooted in the \textit{S-expansion} procedure. This superalgebra, originating in the Soroka-Soroka algebra (with generator $Z_{ab}$ besides $J_{ab}$ and $P_{a}$), represents a much simpler configuration than found in Refs.~\cite{Bonanos:2009wy, Concha:2019icz}). Similar to our previous paper \cite{Durka:2021glk}, this results from extensive computer-assisted validation of the super Jacobi identities among thousands of possible algebra candidates.

We provide Chern-Simons 3D action corresponding to new superalgebra \eqref{resonant_algebra_JPZ+QY+T}. Compared to the known $\mathcal{N}=2$ AdS example, the new CS model provides modified term distribution into sub-invariant sectors. 

The construction of the Chern-Simon action in 3D is straightforward and automatically guarantees the gauge invariance. It would be interesting to perform the construction of the action in 4D. However, in such a case, one deals with the Lorentz invariance, and to assure the supersymmetry, one needs to apply the $1.5$ order formalism. Therefore we leave 4D construction, based on methods already employed in deformed BF theory \cite{Durka:2011gm, Durka:2012wd} and MacDowell-Mansouri models \cite{Gocanin:2019ota, Concha:2014tca, MacDowell:1977jt}, for a future work. Similarly, discussing a broader perspective on the closing of resonant algebras, including the Poincar\'e-like scenarios, will be a subject of separate studies. 

Our general framework explores the generalization of $\mathcal{N}$-extended supergravities, where the extending algebras go through including not only fermionic generators but the additional Lorentz-like and Poincar\'e-like generators. Adding new bosonic generators resembles extending superalgebras to contain a higher number $\mathcal{N}$ of the fermionic generators. But such procedure admittedly still misses direct field/physical interpretation. From the action perspective, we see an appearance of the additional field equations and the nontrivial constraints of systems with respect to the standard AdS case.

Regarding $\mathcal{N}=2$, our goal, besides systematically getting all possible algebra realizations, is to explore the nontrivial breaking of the internal symmetries or soften some of the constraints. Indeed, with more generators and corresponding fields, we may introduce nontrivial modifications at the level of the action or equations of motions.

%%%%%%%%%%%%%%%%%%%%%% Acknowledgments %%%%%%%%%%%%%%%%%%%%%%%%%%%%%%%%%%%%%%%%%%%%%%%%%%%%%%

\section*{Acknowledgments}
We thank prof. Jerzy Lukierski for reading the manuscript and for many valuable insights and comments. We also thank prof. Dragoljub Go\v{c}anin for pointing out helpful literature concerning $\mathcal{N}=2$ Poincar\'e/AdS superalgebras. Additionally, we thank the anonymous reviewer for pointing out the results of \cite{Concha:2020tqx}.

The research project for both authors was partly supported by the program ''Excellence initiative - research university'' for the years 2020-2026 for the University of  Wroc\l{}aw.

%%%%%%%%%%%%%%%%%%%%%%%%%%%%%%% Appendix %%%%%%%%%%%%%%%%%%%%%%%%%%%%%%%%%%%%%%%%%%%%%%%%%%%%%%%

\appendix

%%%%%%%%%%%%%%%%%%%%%%%%%%%%% Appendix A %%%%%%%%%%%%%%%%%%%%%%%%%%%%%%%%%%
\section{3D AdS $\mathcal{N}=2$ superalgebra}
\label{Appendix_A}

For completeness, we provide $JPZ+QY+T$ configuration for the AdS $D=3$ $\mathcal{N}=2$:
\begin{align}\label{The_superalgebra_AdS}
	\left[ J_{a},J_{b}\right] &=\epsilon
	_{abc}J^{c}\,,\qquad \left[ J_{a},P_{b}\right] =\epsilon _{abc}P^{c}\,,\qquad\left[ P_{a},P_{b}\right] =\epsilon _{abc}J^{c}\,,  \notag\\
	\left[J_{a},Q^I_{\alpha }\right] &=\frac{1}{2}\,\left( \Gamma _{a}\right) _{\alpha}^{\beta }Q^I_{\beta }\,,\qquad \left[ P_{a},Q^I_{\alpha }\right]	=\frac{1}{2}\,\left( \Gamma _{a}\right) _{\alpha }^{\beta }Q^I_{\beta}\,,\qquad \left[T,Q^I_{\alpha}\right]=\epsilon^{IJ} Q^J\,,\notag\\ 
	\left\{ Q^I_{\alpha },Q^J_{\beta }\right\} &=-\delta^{IJ}\left(\left( \Gamma^{a} C\right) _{\alpha \beta } P_{a} + (\Gamma ^{a}C )_{\alpha \beta} J_{a} \right)+\epsilon^{IJ}C_{\alpha\beta}T\,.
\end{align} 
The algebraic structure can be highlighted by the convenient table description shown in \cite{Durka:2021glk, Concha:2020atg}, which allows us to focus on the type of the generator outcomes, leaving aside the particular structure constants. For the Majorana supercharges $Q^1\equiv Q$ and $Q^2\equiv Y$ written explicitly, the superalgebra \eqref{The_superalgebra_AdS} in the table form is given by
\begin{small}
	\begin{align}
		\label{table-AdS}
		\begin{array}[t]{c|cc|c}
			[\,,\,] & J & P & T \\ \hline
			J & J & P & 0\\
			P & P & J & 0\\\hline
			T & 0 & 0 & 0
		\end{array}
		\qquad
		\begin{array}[t]{c|cc}
			[\,,\,]	& Q & Y \\ \hline
			J & Q & Y \\
			P & Q & Y \\\hline
			T & Y & Q \\
		\end{array}
		\qquad
		\begin{array}[t]{c|cc}
			\{\,,\,\}	& Q & Y \\ \hline
			Q & P+J & T \\
			Y & T & P+J 
		\end{array}
	\end{align}
\end{small}
We highlight the difference between the AdS \eqref{superalgebra_AdS} and our new $\mathcal{N}=2$ superalgebra, by presenting \eqref{resonant_algebra_JPZ+QY+T} in the table-form: 
\begin{small}
	
	\begin{align}\label{table-resonant}
		\begin{array}[t]{c|ccc|c}
			[\,,\,] & J & P & Z & T \\ \hline
			J & J & P & Z & 0\\
			P & P & Z & P & 0\\
			Z & Z & P & Z & 0\\  \hline
			T & 0 & 0 & 0 & 0
		\end{array}
		\qquad
		\begin{array}[t]{c|cc}
			[\,,\,] 	& Q & Y \\ \hline
			J & Q & Y \\
			P & Q & Y \\
			Z & Q & Y \\\hline
			T & Y & Q
		\end{array}
		\qquad 
		\begin{array}[t]{c|cc}
			\{\,,\,\}	& Q & Y \\ \hline
			Q & P+Z & T \\
			Y & T & P+Z
		\end{array}
	\end{align}
	
\end{small}

%%%%%%%%%%%%%%%%%%%%%%%%%%%%% Appendix B %%%%%%%%%%%%%%%%%%%%%%%%%%%%%%%%%%

\section{3D AdS $\mathcal{N}=2$ supergravity}
\label{Appendix_B}

The connection corresponding to the supersymmetric AdS is given by
\begin{equation}
	\label{connection-AdS}
	\mathbb{A}=\omega^{a}J_{a}+\frac{1}{\ell} e^{a}P_{a}+\frac{1}{\sqrt{\ell}}\psi^{\alpha}Q_{\alpha}+\frac{1}{\sqrt{\ell}}\chi^{\alpha}Y_{\alpha}+\frac{1}{\ell} a\, T\,,
\end{equation}
where we assure proper dimensionality by the introducing length parameter $\ell$ related to the cosmological constant $\Lambda=-1/\ell^2$. 

The super-curvature two-form $\mathbb{F}=d\mathbb{A}+\frac{1}{2}\left[[\mathbb{A},\mathbb{A}\right]]$ is built straightforwardly from $\mathbb{A}$ connection,
\begin{equation}
	\label{curvature-AdS}
	\mathbb{F}=F^{a}J_{a}+\frac{1}{\ell} T^{a}P_{a}+\frac{1}{\sqrt{\ell}}\mathcal{F}^{\alpha}Q_{\alpha}+\frac{1}{\sqrt{\ell}}\mathcal{G}^{\alpha}Y_{\alpha} +\frac{1}{\ell} f(a)\,T\,,
\end{equation}
where
\begin{align}
	F^{a}& =\mathcal{R}^{a}(\omega)+\frac{1}{2\ell^2}\epsilon^{abc} e_b e_c+\frac{1}{2\ell}\,\bar{\psi}\Gamma^{a}\psi+\frac{1}{2\ell}\bar{\chi}\Gamma^{a}\chi\,,  \notag \\
	T^{a}& =D_\omega e^{a}+\frac{1}{2}\bar{\psi}\Gamma^{a}\psi+\frac{1}{2}\bar{\chi}\Gamma ^{a}\chi  \,,  \notag \\
	f& =da-\frac{1}{2}(\bar{\psi}\chi -\bar{\chi}\psi) \,,  \notag \\
	\mathcal{F} & =\mathcal{D}_{\omega }\psi +\frac{1}{2\ell}\,e^{a}\Gamma_{a}\psi-\frac{1}{\ell}\,a\, \chi\,,  \notag\\
	\mathcal{G} & =\mathcal{D}_{\omega }\chi +\frac{1}{2\ell}e^{a}\Gamma_{a}\chi +\frac{1}{\ell}\,a\,\psi \,.
\end{align}
Evaluation of the invariant tensors eventually leads to the corresponding action:
\begin{eqnarray}\label{Action-AdS}
	I_{CS}^{AdS} &=&\frac{k}{4\pi }\int \left[ \alpha _{0}\left(
	\omega ^{a}d\omega _{a}+\frac{1}{3}\epsilon^{abc}\omega _{a}\omega
	_{b}\omega _{c}+\frac{1}{\ell^2}e_{a}D_{\omega }e^{a}+\frac{2}{\ell}\bar{\psi}\mathcal{F}+\frac{2}{\ell}\bar{\chi}\mathcal{G}+\frac{2}{\ell^2}a\,f(a) \right) \right.  \notag \\
	&&\left. +\alpha _{1}\left( \frac{2}{\ell}\mathcal{R}^{a}e_{a}+\frac{1}{3\ell^3}\epsilon
	^{abc}e_{a}e_{b}e_{c}+\frac{2}{\ell}\bar{\psi}\mathcal{F} +\frac{2}{\ell}
	\bar{\chi}\mathcal{G} + \frac{2}{\ell^2} a\, f(a)\right) \right] \,.
\end{eqnarray}
with emergence of the sub-invariant sectors given by the arbitrary constants: $\alpha_0$ and $\alpha_1$. In the first line we recognize ``exotic`` Lagrangian, whereas other is a standard "gravitational" one. 

%%%%%%%%%%%%%%%%%%%%%%%%%%%%% Appendix C %%%%%%%%%%%%%%%%%%%%%%%%%%%%%%%%%%

\section{$\mathcal{N}=2$ super $AdS \oplus Lorentz$ }
\label{Appendix_C}

Let us underline the special role of the Lorentz $J$ generator that introduces the spin connection $\omega$ and preserves all of other generators in the (anti)-commutation relations. Note, however, that the Soroka-Soroka $\mathfrak{C}_4$ algebra can be expressed as the direct sum of the AdS and Lorentz algebras \cite{Durka:2016eun, Soroka:2006aj}. It is done by the change of basis where  the generators are rearranged as follows $L_{AB}=(L_{ab}=Z_{ab},L_{a4}=P_{a})$ and $N_{ab}=J_{ab}-Z_{ab}$. Then  the superalgebra \eqref{resonant_algebra_JPZ+QY+T} reads
\begin{small}
	\begin{align}
		\begin{array}[t]{c|ccc|c}
			[\,,\,] & L_{..} & L_{.} & N_{..} & T \\ \hline
			L_{..} & L_{..} & L_{.} & 0 & 0\\
			L_{.} & L_{.} & L_{..} & 0 & 0\\
			N_{..} & 0 & 0 & N_{..} & 0\\  \hline
			T & 0 & 0 & 0 & 0
		\end{array}
		\qquad
		\begin{array}[t]{c|cc}
			[\,,\,] 	& Q & Y \\ \hline
			L_{..} & Q & Y \\
			L_{.} & Q & Y \\
			N_{..} & 0 & 0 \\\hline
			T & Y & Q
		\end{array}
		\qquad 
		\begin{array}[t]{c|cc}
			\{\,,\,\}	& Q & Y \\ \hline
			Q & L_{.}+L_{..} & T \\
			Y & T & L_{.}+L_{..}
		\end{array}
	\end{align}
\end{small}
Evidently the above tables show that \eqref{resonant_algebra_JPZ+QY+T} represents the direct sum of the super AdS $\mathcal{N}=2$ (spanned by $L_{ab}, P_{a}, Q^I_{\alpha}, T)$ and the Lorentz algebra given by $N_{ab}$.
Note, however, that neither the $L_{ab}$ nor $N_{ab}$ are the physical Lorentz generator as they break general covariance. Although such a basis seems inappropriate to construct physical CS supergravity action, it could be useful for many computational reasons. For instance, one can study asymptotic dynamics. The asymptotic symmetry of $\mathcal{N}=2$ super AdS is given by a $(1,1)$ superconformal structure \cite{Lodato:2016alv}, while the boundary dynamics of $so(2,1)$ CS theory is described by the Virasoro algebra. Combining this with the additional Lorentz generator, one can use it to analyze the general Soroka-Soroka setup.

%%%%%%%%%%%%%%%%%%%%%%%%%%%%%%% Bibliography %%%%%%%%%%%%%%%%%%%%%%%%%%%%%

\normalem
\bibliography{main}
\bibliographystyle{unsrt}
%\bibliographystyle{apsrev4-1}

%%%%%%%%%%%%%%%%%%%%%%%%%%%%%%%%%%%%%%%%%%%%%%%%%%%%%%%%%%%%%%%%%%%%%%%%%%%
\end{document}